\documentclass[journal]{IEEEtran}
\IEEEoverridecommandlockouts
\usepackage{cite}
\usepackage{amsmath,amssymb,amsfonts}
\usepackage{algorithm}
\usepackage{algorithmicx}
\usepackage{graphicx}
\usepackage{textcomp}
\usepackage{xcolor}
\usepackage[justification=centering]{caption}
\usepackage{algpseudocode	}
\makeatletter
\newcommand{\removelatexerror}{\let\@latex@error\@gobble}
\makeatother
\usepackage{comment}
\usepackage{booktabs}

\usepackage{soul}
\soulregister\cite7 
\soulregister\citep7 
\soulregister\citet7 
\soulregister\ref7
\soulregister\pageref7

\def\BibTeX{{\rm B\kern-.05em{\sc i\kern-.025em b}\kern-.08em
    T\kern-.1667em\lower.7ex\hbox{E}\kern-.125emX}}
\begin{document}

\title{A Wearable ECG Monitor for Deep Learning Based Real-Time Cardiovascular Disease Detection}

\author{%
\begin{tabular}{c}
Peng Wang, Zihuai Lin, Xucun Yan, Zijiao Chen, Ming Ding, Yang Song, and Lu Meng\tabularnewline
\tabularnewline
\end{tabular}
\thanks{
Peng Wang, Zihuai Lin, Xucun Yan and Zijiao Chen are with the School of Electrical
and Information Engineering, The University of Sydney, Australia (e-mail:wpwd1986@hotmail.com,
zihuai.lin@sydney.edu.au, xucun.yan@sydney.edu.au and zche8039@uni.sydney.edu.au).

Ming Ding is with Data61, CSIRO, Australia (e-mail: ming.ding@data61.csiro.au).

Yang Song is with Beijing Irealcare Pte Ltd, P. R. China (e-mail: syang@irealcare.com).

Lu Meng is with Northeastern University, P. R. China (e-mail: menglu1982@gmail.com).
}}


\maketitle

\begin{abstract}
Cardiovascular disease has become one of the most significant threats endangering human life and health. Recently, Electrocardiogram (ECG) monitoring has been transformed into remote cardiac monitoring by Holter surveillance. However, the widely used Holter can bring a great deal of discomfort and inconvenience to the individuals who carry them. We developed a new wireless ECG patch in this work and applied a deep learning framework based on the Convolutional Neural Network (CNN) and Long Short-term Memory (LSTM) models. However, we find that the models using the existing techniques are not able to differentiate two main heartbeat types (Supraventricular premature beat and Atrial fibrillation) in our newly obtained dataset,  resulting in low accuracy of 58.0\,\%. We proposed a semi-supervised 
method to process the badly labelled data samples with using the confidence-level-based training. 
The experiment results conclude that the proposed method can approach an average accuracy of 90.2\,\%, i.e., 5.4\,\% higher than the accuracy of conventional ECG classification methods.


\end{abstract}

\begin{IEEEkeywords}
Cardiovascular disease; ECG monitoring; Holter; ECG-IoT; AI data analysis.
\end{IEEEkeywords}



\maketitle
\section{Introduction}
\label{sec:introduction}
Recently, Internet of Things (IoT) has captured much attention in academics and industries. 
Generally speaking, 
IoT can be viewed as an extension of the Web application, 
which is the Internet connection with real-life things, 
instead of information.
IoT is an essential part of the new generation of information technology~\cite{d8}, 
which can be widely used in the fusion of the network through communication technologies
such as smart perception, recognition technology, and pervasive computing~\cite{d2}. 
At the same time, 
the application of IoT in medical treatment has also become increasingly popular~\cite{d8}, 
giving birth to smart medicare, 
e.g. visualization of material management, 
digitization of medical information, 
and digitization of therapeutic process.




Smart healthcare is defined by the technology that leads to better diagnostic tools, 
better treatment for patients, 
and devices that can improve the quality of life for everyone. 
In this paper, 
we focus on smart healthcare concerning cardiovascular diseases~\cite{d5}. 
Cardiovascular disease is one of the leading causes of death worldwide.
The number of people who die of cardiovascular disease annually is estimated to be 17.7 million,
which accounts for about 31\,\% of the global death toll. 
Some heart diseases are asymptomic and undetectable such as atrial fibrillation, 
which needs to monitor ambulatory electrocardiography in a long term.

Electrocardiogram (ECG) is a kind of diagram generated from electrical activities of heart in terms of voltage and time. 
It can be recorded by some monitoring devices. 
The existing ECG monitoring devices typically require a wearing of 24-48 hours. 
During such a period, 
the devices collect the wearer's heart related signals. 
Hospitals then analyze the signals and diagnose the wearer's cardiovascular conditions if any. 
This post-analysis of the heart signals, 
although can often identify heart abnormality given sufficient data, 
lacks the ability of real time prediction of the wearer's cardiovascular conditions.
As such, 
it is generally very hard to provide critical warning to the wearer in real time.
In addition, 
such ECG monitoring devices are often bulky and impractical for a person to wear all the time. 
Fortunately, 
digital health innovation (DHI) driven by data science and artificial intelligence (AI) promise smart healthcare. 

In this paper, 
we propose a wearable ECG monitoring system, IREALCARE, 
comprised of an integrated ECG sensor for continuous, 
long-term remote ECG monitoring, 
and an AI platform for abnormal ECG patterns recognition. 
The collected new dataset with IREALCARE is classified with the existing Convolutional Neural Network (CNN)~\cite{d42} and Long Short-term Memory (LSTM)~\cite{d43}. 
However, 
it can be found that these models were inadequate for classification of data from IREALCARE with high accuracy. 
It is because that the dataset collected by the patch is much noisier than the Holter due to the limited resource inside the IoT  ECG patch,  
which  could  generate  a  significant number  of  Electromagnetic  Interference.  
In  addition,  
a  considerable amount of movement interference will be introduced when the patients wear the patch every day. 
Besides, 
there might also be some errors in labelling as well. 
Labelling such a large dataset tends to be erroneous due to human mistakes and medical misjudgment. 
Hence, we propose a ResNet with confidence Level based training to solve this issue. 
 
The main contributions of this paper are listed below:

\begin{itemize}
\item A novel portable wireless real-time single lead ECG patch was developed for ECG monitoring.
\item A low-complexity semi-supervised training scheme based on a deep residual block with confident training was proposed to robustly classify the ECG  signals  that are collected  under  an  imperfect condition.
\item Comparisons between the proposed scheme and existing models were  made. The proposed scheme can produce a excellent accuracy of 90.2\,\%, 
i.e., 5.4\,\% higher than the accuracy of conventional ECG classification methods.
\end{itemize}

The overall structure of this paper is outlined as follows. 
Section II introduces the background of cardiovascular diseases and the state-of-the-art approach for 
cardiovascular disease detection. 
The structure of the proposed Electrocardiogram (ECG)-IoT system is described in Section III. 
In Section IV, 
we discuss the features and advantages of the proposed smart ECG-IoT system. 
In Section V, 
we present a reliability analysis and comparison of the experimental results.
Conclusions and recommendations are given in Section VI.

\section{Related Work\label{related work}}
\subsection{ECG Monitoring System}
Cardiovascular disease is one of the most life-threatening diseases for human being, 
and the incidence, mortality and disability of global cardiovascular events have been increasing in recent decades~\cite{d12}.
Malignant arrhythmia is one of the major causes of sudden cardiac syncope or sudden cardiac death outside the hospital~\cite{d13}.
An important basis for the current diagnosis of arrhythmia is the heart rate captured by the ECG signals of the patient. 

Traditional ECG measuring machines are mostly of bulk sizes and stationary~\cite{d5}. 
People need to go to qualified institutions or hospitals for measurement,
and they need to go through complicated procedures such as registration, queuing,
and payment to obtain measurement results \cite{d14}.
This is both time-consuming and expensive. 
More importantly, 
the conventional ECG measurement is conducted by the patient in a supine position,
and thus the measurement time is short,
and only a small amount of data about the heart state can be obtained \cite{d8,d5,d12,d13,d14,d17}.
Therefore, 
if the heart rate is abnormal only within a limited time period, 
the probability of detecting such anomaly could be low. 

Fortunately, 
with the development of mobile monitoring technology and information transmission technology, 
portable ECG measurement devices becomes feasible.
However, 
most of the portable ECG devices including the commonly used Holter are still using multi-channel ECG sensors,
requiring multiple electrodes to collect ECG data and most portable ECG electrodes are wired to a central control unit, 
which brings a great deal of discomfort to the people who carry them \cite{d5,d14}

In order to reduce the burden on the wearer and dynamically monitor the patient's heart rate condition for a long time,
a new wireless smart ECG patch, 
which is named IREALCARE has been designed in this paper. 
This patch can dynamically collect wearer\textquoteright s ECG signal in real time and the battery can support up to 72 active hours.  
The collected data via the IREALCARE ECG patch can be sent to a smart phone APP and then to the cloud for further data analysis. 
Through big data analysis in the cloud, 
the system can diagnose the wearer's health condition. 
The data and results can also be accessed and verified by a medical staff. 
The authorized results can be sent to the wearers' smart devices, 
such as mobile phones, IPAD or computers. 

\subsection{ECG Signal Classification}
\begin{figure}[htb]
  \centering   
  \includegraphics[width=3in]{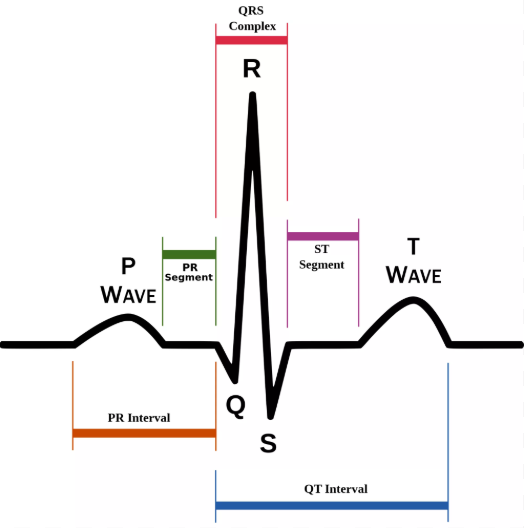}
  \caption{The composition of each wave and wave band of ECG}
\end{figure}

ECG is an objective indicator of the occurrence, transmission, and recovery of cardiac arousal \cite{d23}. 
ECG provides essential reference information to the primary function of the heart and its case study. 
It can be used for all kinds of examinations on heart rate disorders or heart-related disorders,
but also a vital basis to determine cardiovascular disease. 
As shown in Fig. 1, 
there are three waveforms, two periods and two typical segments of ECG \cite{d25}.

The three waveforms are (1) P waves produced by atrial agitation, 
the average p-wave time length is $0.12$ second, 
and the average peak is $0.25mv$. 
\begin{table}[htb]
  \centering
  \renewcommand\arraystretch{1.5}
  \begin{tabular}{|c|c|c|}
    \hline
    ECG Waveform & \textbf{Heart Electrical Acitvity} & \textbf{Standard Value} \\
    \hline
    P Wave & Atrial removal of the pole & 0.06-0.12 \\
    \hline
    T Wave & Ventricular complex polarization & 0.1-0.25 \\
    \hline
    QRS Complex & Ventricular removal pole & 0.06-0.11 \\
    \hline
    PR Interval & Chamber conduction time & 0.12-0.2 \\
    \hline
    {2}{*}{QT Interval} & Time of ventricular depolarization & 
    {2}{*}{(0.36-0.44)} \\
     & to complete repolarization & \\
    \hline
    ST Interval & Ventricular removal is complete & 0.12-0.16 \\
    \hline
  \end{tabular}
  \caption{ECG standard Value of normal people}
\end{table} 
When the atrium expands, 
the P wave can exhibit a higher tip or double-peak waves when the atria are abnormal. 
(2) The QRS wave group is the reaction of the point of the left and right ventricle. 
The QRS wave group represents the ventricular depolarization, 
and the agitation time length of normal duration should be less than $0.11$ seconds. 
When the conduction block, 
ventricular enlargement or hypertrophy of the left and right bundle of the heart appears, 
the QRS wave group would be enlarged, deformed and prolonged. 
(3) T wave represents the potential of the recovery after ventricular agitation. 
The change of T wave is affected by a variety of factors, 
and normal T wave should be in the same direction as the QRS main wave.

There are two kinds of intervals: 
(1) PR interval:
a regular PR interval is between $0.12$ seconds and $0.20$ seconds. 
When the conduction from atrial to ventricular is blocked, 
the PR interval will be prolonged, 
or the ventricular wave disappears after P wave. 
(2) Q-T period:
The time duration between the start from the QRS wave group and the end of T wave. 
It represents the whole process of the cardiac depolarization and repolarization. 
Usually, 
the time between the Q-T period should not be greater than $0.44$ seconds.
There are two typical wave segments: 
(1) The PR segment starts from the second half of the P wave to the beginning of the QRS wave group.
This segment of healthy people is close to the baseline, 
and the distance between the segment and the baseline is generally no more than $0.05mm$. 
(2) S-T segment: 
From the end of the QRS wave group to the beginning of the T wave. 
Similarly, 
the S-T segment of a healthy person is close to the baseline. 
Typical ECG values for a healthy person are shown in Table I.

Researches mainly put their attentions on ECG signal preprocessing and classification. The objective of preprocessing signals is to extract useful features by suppressing noises and amplifying desired characteristics. The authors of \cite{d30} used Wavelet Transform (WT) to detect QRS complex. The algorithms to automatically detect ECG segments was proposed in \cite{d31,Yeh09C}. Adaptive Least mean squares (LMS) filtering and adaptive Recursive least squares (RLS) filtering was explored by Martinek in order to find optimal parameters for filters \cite{d32}. Some researchers tended to rely on certificated cardiologists who can manually select QRS points and label heart diseases. Although this screening approach requires large human resources and cost, it is the most reliable method among aforementioned ways.

Traditional classifications identify features through Random Forest (RF), Support Vector Machine (SVM) and k-Nearest Neighbors (k-NN), which are suitable for small data sets. Zabihi et. al used Random Forest to detect atrial fibrillation \cite{d34} and Alickovic et. al applied same algorithm for medical decision \cite{d33}. Raj et. al proposed SVM based method to classify the ECG signal \cite{Raj17E}. The comparison between SVM and k-NN on ECG was made by Diker in 2018 \cite{d35}. For large data sets, normally, it is more general to use deep leaning methods such as Artificial neural networks (ANN) \cite{d36}, Convolutional Neural Network \cite{d37} and Recurrent Neural Network (RNN) \cite{d38}. ANNs are widespread and efficient tools for pattern recognition problems like ECG heartbeat classification. 
Many techniques have been introduced which are based on different NN structures \cite{d26}. 
NN structure is famous according to its fast learning algorithms such as a radial basis function-based neural network (RBFNN). RBFNNs are nonlinear hybrid networks which are exactly containing a single hidden layer of neurons. The input layer forwards the coordinates of the input vector to each of the nodes in the hidden layer. Every node in the hidden layer then computes an activation according to the associated radial basis function (RBF). At the end, each node in the output layer computes a linear combination of the results of the activation functions of the hidden nodes \cite{d27}. Deep learning has drawn much attentions because its high capabilities of data processing and prediction accuracy. This is due to the injunction of great amounts of data, providing massive nonlinear combinations to find an optimal solution.

However, 
these researches were all built on the MIT-BIH dataset, 
which is a very classic public and wholesome dataset~\cite{d39}. 
The data is clean and collected from a 24-hour monitoring Holter~\cite{d41}, 
which has a significant difference from the data collected by an ambulatory IoT patch. 
In many applications involving ECG, 
the collected data will not be as clean and high-quality as the MIT-BIH dataset and therefore requires extra data analysis.

\section{\label{sec:Preliminary-1}The Architecture of the proposed ECG Monitoring System}

In this section, 
we present the architecture of the proposed ECG monitoring system. 
\subsection{ECG Monitoring Sensor}


\begin{figure}[htb]
\centering
\includegraphics[width=3in]{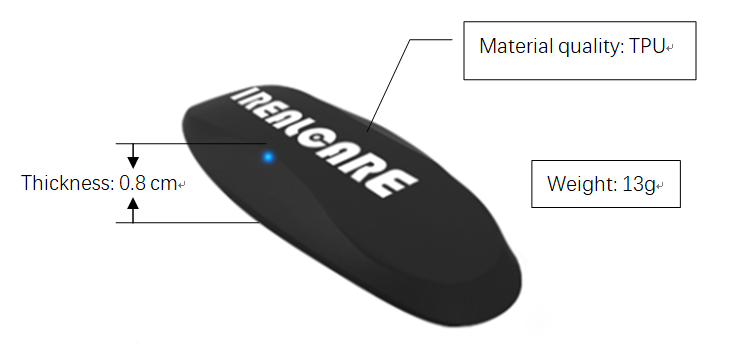}
\caption{Outlook of the IREALCARE ECG Monitoring Sensor}
\label{fig2}
\end{figure}

An IREALCARE sensor is shown in Fig. \ref{fig2}. 
The sensor is small and lightweight with only 13 grams. 
The soft and comfortable shell of IREALCARE is made of high polymer material--Thermoplastic polyurethane (TPU). 
The good plasticity of TPU makes the sensor compact, 
and the overall equipment thickness is only 0.8cm. 
The material also possesses desirable properties such as resistance to wear, water, mildew, and coldness. 

IREALCARE's excellent portability differentiates itself from the traditional bulk-size instrumentation, 
and its long-term continuous monitoring ECG capability enables monitoring patients who are in need of year-round healthcare.

IREALCRE is a single-channel dynamic ECG instrument, 
which can record ECG signals continuously for 72 hours. 
The long time wearing can increase the probability of detecting the symptoms of arrhythmia 
which are not easily detectable in routine electrocardiogram examination. 
Therefore, 
it can provide an important objective basis for clinical diagnosis, treating, and judging the curative effect. 
Currently used ECG monitors in clinic facilities are the 12-lead full-record devices (Holter) \cite{d5}. 
In contrast, IREALCARE is more comfortable for wearing \cite{d23}. 

\subsection{Overview}

\begin{figure*}[t]
\centering
\includegraphics[width=0.70\textwidth]{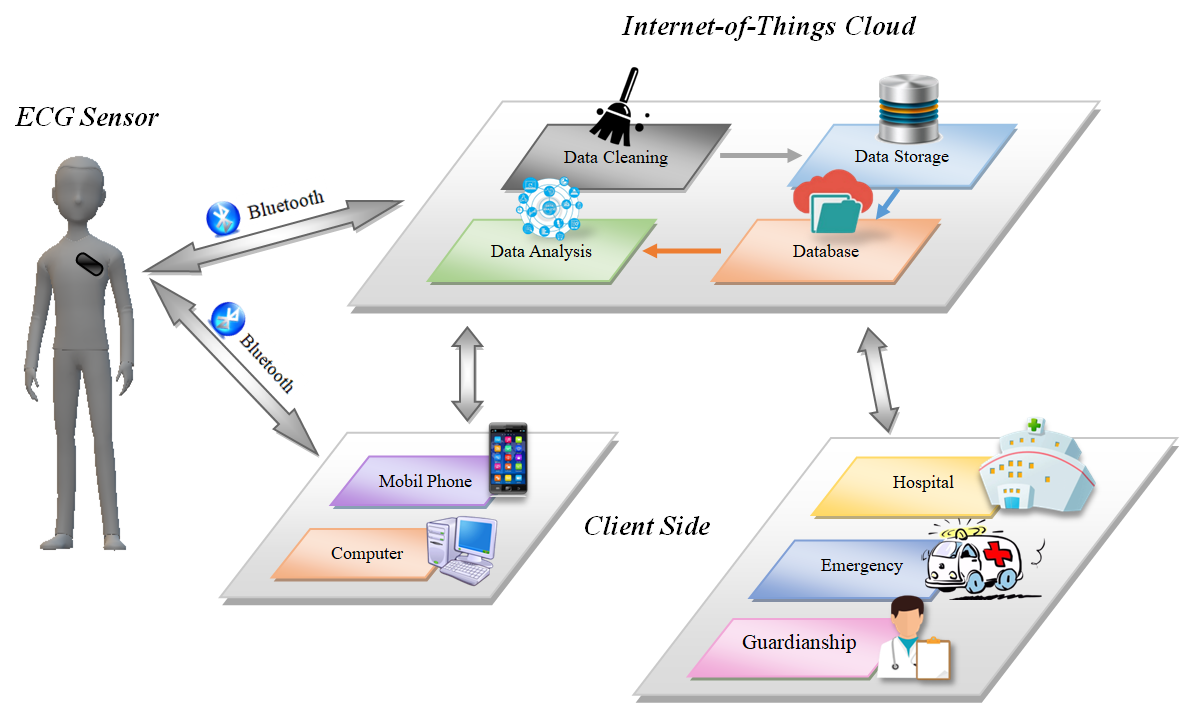}
\caption{The architecture of the ECG monitoring system}
\label{fig3}
\end{figure*}

This proposed ECG monitoring system mainly uses the IoT technology to achieve real-time monitoring of ECG signals, 
analyze the data and establish a platform for remote communications between doctors and patients, 
which can perform prevention and treatment of cardiovascular disease~\cite{d4}. 
As shown in Fig. \ref{fig3}, 
the proposed system consists of three components: 
ECG sensing, IoT cloud, and client.

\begin{figure}[h]
\centering
\includegraphics[width=3in]{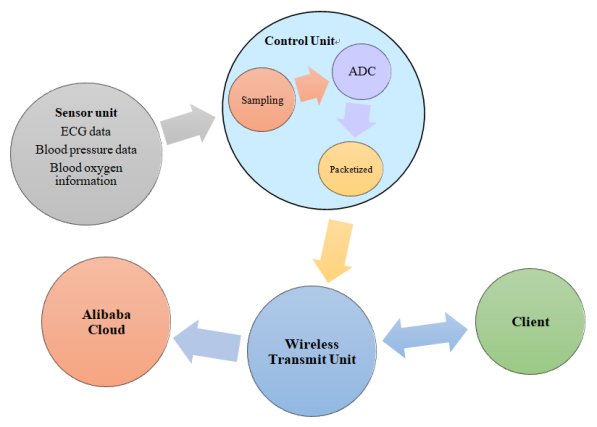}
\caption{Structure of the ECG sensor}
\label{fig4}
\end{figure}

An ECG sensor can generate an electrocardiogram by collecting electrical signals generated by each heartbeat of a human body. 
Fig. \ref{fig4} illustrates three main units of a sensor: 
a sensing unit, 
a control unit, 
and a wireless transmit unit. 
\begin{itemize}
    \item In the sensing unit, 
    the sensor collects ECG data and then transmits the data to the control unit. 
    \item The signals obtained in the control unit are sampled and processed at a sampling frequency of 250Hz. 
    The sampled electrical signals are then converted to digital signals and ready to be sent to the wireless transmit unit. 
    The 24-bit analog-to-digital converter (ADC) is used in our system implementation. 
    \item 
    The wireless transmit unit interacts with the mobile device on the client side via Bluetooth. 
\end{itemize}

The ECG sensor designed in this paper also has a charging unit with a rechargeable lithium battery, 
which can last for a maximum of 72 hours without recharging. 
In this way, 
the heart rate status of the wearer can be continuously monitored for a long time to achieve real-time detection of cardiovascular diseases such as cardiac arrhythmia.

In the IoT cloud, 
the received data is first cleaned up, 
and then pre-processed to reduce noise. 
The cleaned and noise-suppressed data will be stored in a memory bank for statistical data analysis and future data extraction. 


Based on a large amount of labeled ECG signals, 
a model can be trained to classify the labeled ECG signals into various heart conditions in the IoT cloud. 

Then, 
we can use the trained model to estimate the classes of new ECG signals, 
i.e. the inference computation. 
The outcome of the inference data analysis in the cloud will be transferred to the client. 
At the client side, 
the data is presented to the user graphically, 
enabling data visualization and making it easier for the wearer to understand the results. 
In this system, 
users not only refer to the wearer, 
but also to the wearer's guardian and medical staff. 
Doctors can retrieve the patients' data from the cloud, 
analyze the original data, 
and the outcome of ECG signal classification, 
to better diagnose the patient's health condition. 
At the same time, 
doctors can provide appropriate treatment recommendations based on the analytical results and send them to the cloud, 
so that patients can get these instructions from the cloud directly instead of going to the hospital.

Furthermore, 
doctors can send some efficacy data to the cloud based on patients' medication, 
thus establishing a database of certain drugs to analyze the effect of the drug. 
In this way, 
patients can not only read the ECG data, classification outcomes and analysis results from the cloud, 
but also get the doctor's professional advice. 
Hence, 
the direct and timely communication can be established between patients and doctors.

\section{\label{sec:Pre-processing}Pre-processing}
The procedure of pre-processing is described as follows. The ECG data was normalized initially and then enlarged by applying sliding window. Finally, the enlarged data is denoised with multiresolution analysis - discrete wavelet transform (DWT). 

\subsection{Enlarging data}

\begin{figure}[htb]
  \centering
  \includegraphics[width=3in]{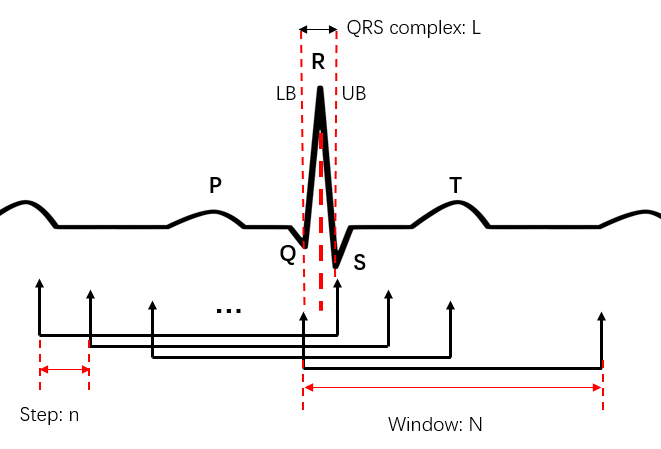}
  \caption{Illustration of sliding process}
  \label{fig5}
\end{figure}

The window sliding method enlarges data to complement less amount of heart disease types so that we can solve the imbalanced data issue. 
The window should always contain at least QRS complex segment, which can be demonstrated in Fig. \ref{fig5}, where the length of QRS complex segment L, window size N and step size n. 
The absolute start point of the last window must be less than or equal to lower boundary (LB) while the absolute end point of the first window must be greater than or equal to upper boundary (UB). In this case, we set N to 600 and n to 20. The number of windows for each type highly depends on the length of QRS complex segment. After this operation, our data is dramatically enlarged.
\subsection{Denoising}
DWT is widely used in signal processing due to its robust and efficient ability of capturing information in both time and frequency domain. Its decomposition formula for signal $s(t)$ can be expressed as 

\begin{equation}
\label{eq1}
s(t) = \sum_{k=-\infty}^{\infty} c_k\phi(t-k)+\sum_{j=-\infty}^{\infty}\sum_{k=-\infty}^{\infty}d_{j,k}\psi(2^jt-k),
\end{equation}
where $\phi$ and $\psi$ are the respective scaling function and wavelet function, and their corresponding coefficients $c_k$ and $d_{j,k}$ are named scaling and wavelet coefficients, respectively \cite{d47}.
Moreover, the first term and the second term of Eq.~\ref{eq1} represent the trend and the local details of $s(t)$, respectively. 
The term $d_{j,k}\psi(2^jt-k)$ is the local residual error between signal approximations at scales $j$ and $j-1$.

\begin{figure}[htb]
  \centering
  \includegraphics[width=3in]{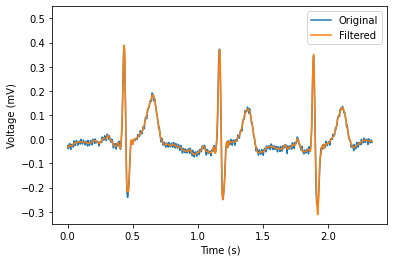}
  \caption{Original signal vs Filtered signal}
  \label{fig6}
\end{figure}

To effectively achieve the functionality of aforementioned two terms, low-pass and high-pass filters are convoluted with signal $s(t)$ for keeping the approximation and detail, generating the shape approximating scale coefficients (CA) and the detailed wavelet coefficients (CD). 
The denosing process can be implemented by removing some detail signals at typical scales. 
In our study, Daubechies wavelet is chosen as mother wavelet $\psi$ and by setting scale number $j$ to 4, five  coefficients (CD1 - CD4 and CA4) are generated. Then, the detail coefficients at level 1 and 2  (CD1 and CD2) are zeroed to remove noises, and the signal was reconstructed with the same mother wavelet and modified coefficients. Fig.~\ref{fig6} shows the signal before and after denoising process. Clearly, the denoised signal maintains the original trend and removes redundant local fluctuations.

\section{\label{sec:Simulation-Result}Results and Analysis}

In this section, 
we present our results and analysis.

\subsection{Data}
The ECG data records are gathered from 62 patients wearing IREALCARE sensors introduced in the previous sections. 
The sample frequency of the ECG data is $300 Hz$.

Each ECG segment in the training set has the length of $1$ second and can contain only one rhythm type.
Several clinical ECG experts with extensive knowledge and experience in arrhythmia detection will label the ECG signal into one of the 5 rhythm classes.
The labelled dataset will be divided into a training and testing set with a ratio of $80\%/20\%$.

In this paper, $3$ heart arrhythmias are selected, together with the additional normal rhythm and unclassifiable heartbeat, there are $5$ output classes in total. We associate the
annotations in this dataset with the beat categories in Association for the Advancement
of Medical Instrumentation (AAMI) EC57 standard \cite{d29}. See
the first two columns of Table II for a detail of mappings.
We selected 227,680 heartbeats (exact 45,536 from each class) to train this model and tested this ECG signal classifier on
45,535 heartbeats (exact 9107 from each class) that are not used
in the network training phase with the consideration of balance in the number of beats in each
category.  
Table~\ref{tab_2} shows the number of $5$ types of ECG signals.

\begin{table}[ht]
\caption{Number for six types of ECG signals}
	\begin{center}
		\begin{tabular}{|c|c|c|}
			\hline
			AAMI EC57[36] &ECG Beat Types&number\\
			\hline
			N&Normal Heartbeat& 3,501,003\\
			V&Ventricular premature beat&	194,469\\
			S&	Supraventricular premature beat&45,536\\
			A&	Atrial fibrillation&96,052\\
			Q&	Unclassifiable beat& 148,020\\
			\hline
		\end{tabular}
	\end{center}
	\label{tab_2}
\end{table}

\subsection{Proposed ECG Classification Approach}
In this section, we propose an ECG classification approach.
To implement this training scheme, we need to find the noisy signals by performing Confidence-level-based Learning.

\subsubsection{Confidence Level based Training}
To improve the training process, we introduce a term called $“$confidence level$”$.
The definition of confidence level is the output probability of the last layer in the model. 
We set a threshold for the confidence level: 
if the data$'$s confidence level is higher than the threshold, 
it is considered as clean data. 
In a confidence level based training scheme, 
to avoid impairing the model, 
only clean data will be trained.

\subsubsection{Residual Network and Residual Block}

\begin{figure}[ht]
\centering
\includegraphics[width=5cm]{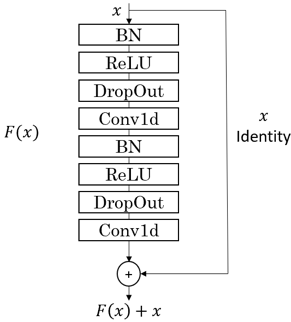}
\caption{Residual Block}
\label{fig_7}
\end{figure}
\begin{figure*}[ht]
\centering
\includegraphics[width=18cm]{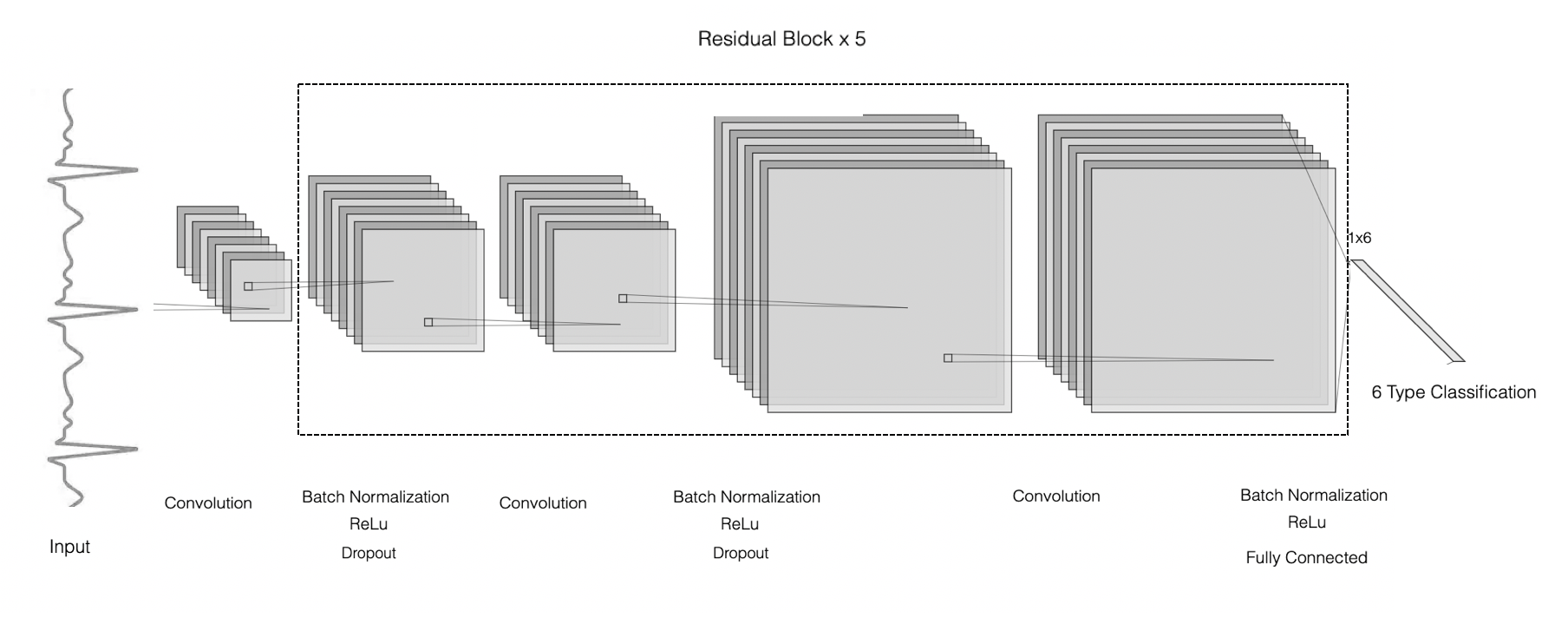}
\caption{Network Structure of ResNet}
\label{fig8}
\end{figure*}
Here, we use the Residual Network (ResNet) as the training model. Residual block is the building block of our network backbone, which consists of two convolutional layers and an identity connection from input to output. The identity connection is the most important part of the residual block. It passes more information from the shallower layer to the deeper layer by a simple addition operation, making a deeper learning network possible. 

Fig. \ref{fig_7} shows the structure of our proposed residual block. In our implementation, a max pooling layer is added in every two residual blocks. And other regulation operations like dropout and batch normalization are also added in the residual block to improve the convergence. 

\subsubsection{Neural Network Backbone Model}

Our model consists of six residual blocks with filter number doubles for each block as shown in Fig. \ref{fig8}. 

The architecture and parameters of the model are listed below: \\

      \begin{tabular}{c}
       \hline \hline
      Model Architecture\\
       \hline 
      1 Conv + 5 Residual Blocks\\
      Residual Block: Batch Normalization + Conv + ReLu\\
       \hline \hline
       1D Conv layers\\
       \hline
      filter size: 3\\
      number filter: 64 to 256 (doubles every 2 residual blocks)\\
      stride=1, padding = 1\\
      initialized without bias\\
       \hline \hline
       Dropout Rate\\
       \hline
       training: 0.5\\
       prediction: 1\\
       \hline \hline
       Activation: ReLu\\
       \hline \hline    
     
\end{tabular}

~\\

\begin{center}
      The parameters of the training processing: \\
      \begin{tabular}{c|c}
       \hline
      \multicolumn{2}{c}{Training Process Parameters} \\
       \hline
      Loss function & Cross Entropy\\
      Optimizer & Adam\\
      Learning Rate & 0.002 (decrease on plateau)\\
      Training epoch & 100\\
      Batch size & 128 with shuffle \\ 
       \hline 
      \end{tabular}	
\end{center}

As input, we provided zero
mean unit variance raw ECG signal segment. Since CNNs require
a fixed window size, we truncated these segments to the
first minute. We also tested other variations of the proposed model by changing the depth, reducing the number
of filters at each layer.


\subsection{Performance metrics}
There are numerous performance measures concerning classification results in general.
Two metrics take the first place in literature which are:
\begin{enumerate}
\item Accuracy, representing the percentage of how much correctly classified beats through all the beats considered (independently of the classes they fit to).
\item Precision, known as positive predictive value (PPV), is the fraction of retrieved instances that are relevant.
\end{enumerate}

\subsection{Experimental results}

We will explain the training procedure with details as below: 
\begin{enumerate}
\item We firstly train the raw data with a classical CNN. 
\item Given the obtained confidence level, we can distinguish the data with perfect label from the data with imperfect label. 
\item The data with good labels will be selected to train the ResNet model.
\end{enumerate}

In Table \ref{tabIII}, we can observe the importance of the optimal value of confident level. We have conducted experiments on our data set with different confident level and obtain the optimal accuracy with confidence level equals to $80\%$. 
The higher the confident level is, the less data would be selected as data with good labels to train the model. The lower the confident level is, the more likely the data has imperfect label. Hence, searching for the optimal confident level is the most important task for impertect labelled data training problem. Once the data and the corresponding labels change, the optimal value changes as well.

\begin{table}[ht]
	\caption{Accuracy of confidence based ResNet with different confidence level}
	\begin{center}
		\begin{tabular}{|c|c|}
			\hline
			Confidence Level&Accuracy \\
			\hline
			99\%&~82.2\%\\
			90\%&~86.5\%\\
			80\%&~89.9\%\\
			70\%&~87.6\%\\
			60\%&~86.4\%\\
			50\%&~84.3\%\\
			40\%&~82.3\%\\
			30\%&~82.3\%\\
			\hline
		\end{tabular}
	\end{center}
	\label{tabIII}
\end{table}

\begin{figure}[ht]
\centering
\includegraphics[width=9cm]{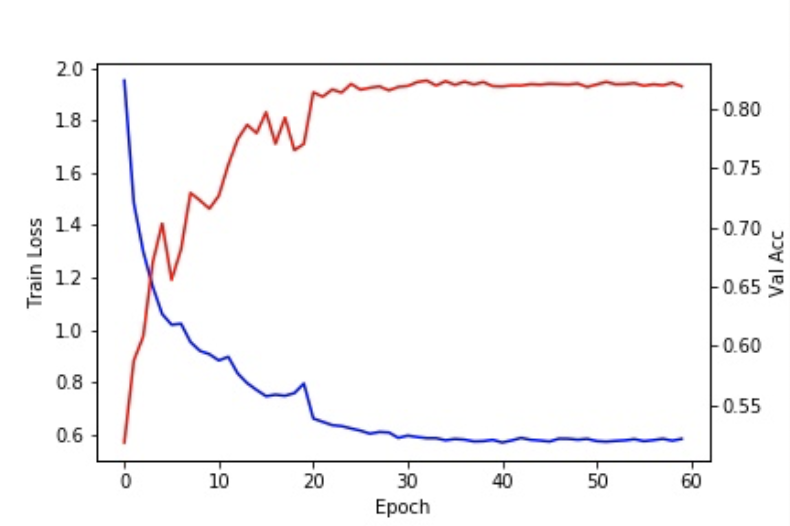}
\caption{Training Accuracy and Loss Plot of 80\% Confidence Level}
\label{fig11}
\end{figure}

It can be observed that the highest accuracy $89.9\%$ can be obtained with the confident level equals to 80\%. 
The accuracy and loss plot of training with $80\%$ of confidence level is shown in \ref{fig11}.

To demonstrate the superiority of the proposed model,
we utilised a number of existing advanced classifiers~\cite{d42,d43,d44,d45,d46} to classify our data. The classification results are compared in Table \ref{table_comparison}.

\begin{table}[ht]
	\caption{Comparison of classification accuracy between different models}
	\begin{center}
		\begin{tabular}{|c|c|}
			\hline 
            Model & Accuracy\\
            \hline
            Referenced CNN \cite{d42} & 58.0\% \\
            Referenced LSTM \cite{d43} & 69.5\% \\ 
            AlexNet \cite{d44} & 70.1\% \\
            VGG16 \cite{d45} &  82.9\%\\
            WBCNN \cite{d46} & 77.7\% \\
            \textbf{Confident-ResNet}  & \textbf{90.2\%} \\
            \hline
		\end{tabular}
	\end{center}
	\label{table_comparison}
\end{table}

\begin{figure}[ht]
\centering
\includegraphics[width=10cm]{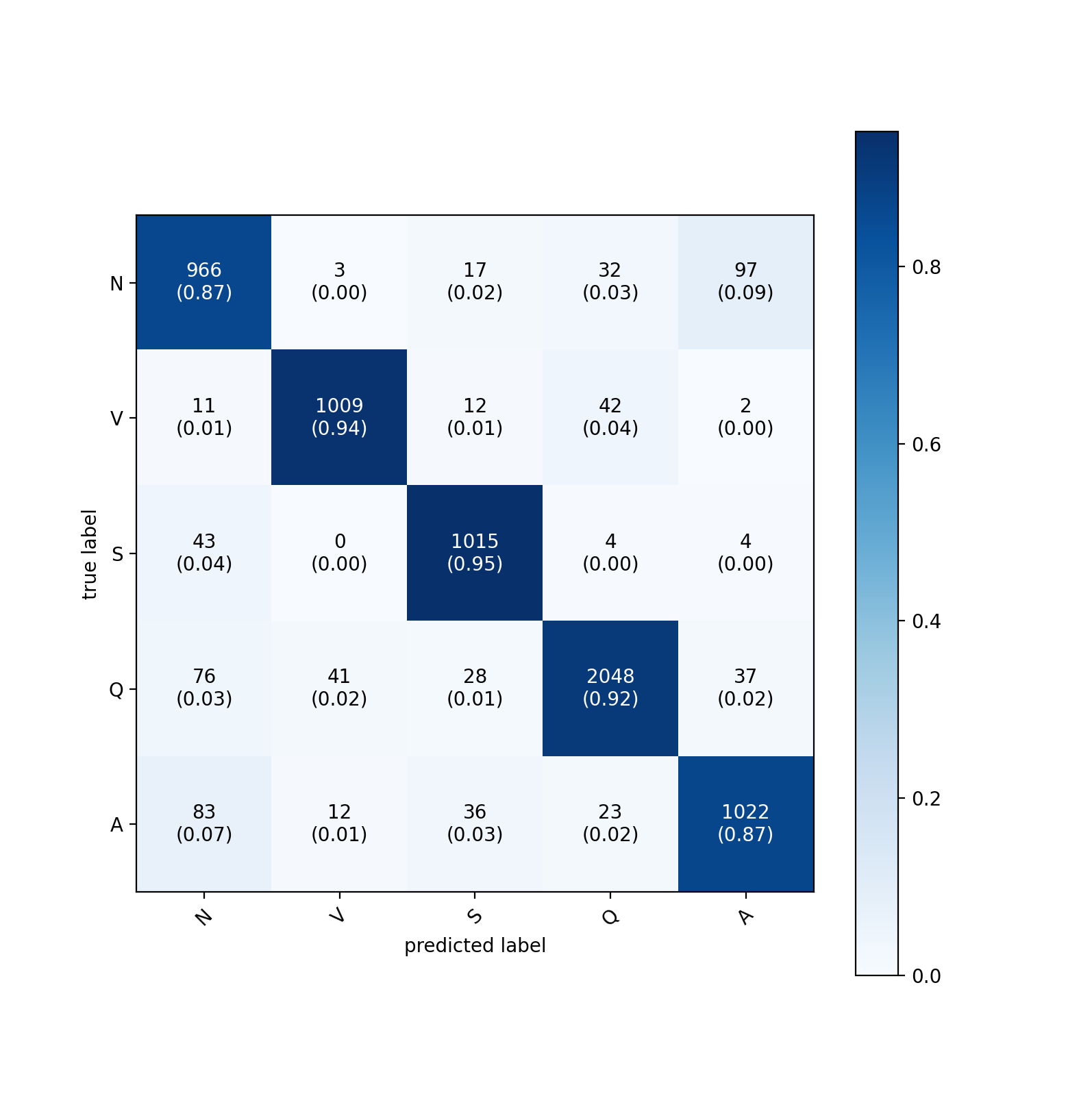}
\caption{Confusion Matrix of Negative ResNet}
\label{fig13}

\end{figure}

Fig. \ref{fig13} demonstrates the confusion matrix of applying the
proposed classifier on the test set. As it can be seen from this figure,
the model can produce accurate estimations. Table \ref{tab5} shows the accuracy and precision obtained on the test beats by feeding the proposed Confident-ResNet. In general, as can be observed from Table \ref{tab5}, these results
show the superiority of the proposed approach achieving the overall classification accuracy for all records of 90.2\%. which is $3-10\%$ better than the state-of-the-art classification method. 
In other words, the Confident-based learning method is more suitable for the badly labelled dataset than the exsiting classification techniques.

\begin{table}[ht]
\caption{Precision and Recall for Negative ResNet}
	\begin{center}
		\begin{tabular}{|c|c|c|c|c|c|}
			\hline
			  &N&V&S&Q&A\\
			\hline
		    mean Precision&0.82 & 0.95& 0.92& 0.95& 0.88\\
			mean Accuracy &0.86& 0.94 &0.95& 0.92& 0.87\\     
			\hline
		\end{tabular}
	\end{center}
	\label{tab5}
\end{table}

\section{\label{sec:Conclusion}Conclusion}

In this paper we provided a new wireless ECG IoT monitoring system and introduced a novel wearable ECG sensor IREALCARE. 
To classify and analyse the acquired ECG signals which do not have perfect labelling, a residual network with newly designed confident-level based training was developed to generate a sophisticated model based on this dataset. 
The experiment results conclude that the proposed method can approach an average accuracy of 90.2\,\%, i.e., 5.4\,\% higher than the accuracy of conventional ECG classification methods. In the future work, we plan to extract more effective features from ECG signals obtained by IREALCARE ECG devices to diagnose heart beat conditions and explore other more suitable classifiers for this ECG dataset classification.

\bibliographystyle{IEEEtran}
\bibliography{Journal_main}


\end{document}